\documentstyle[amssymb,prl,aps,twocolumn]{revtex}

\begin{document}
\draft
\twocolumn[\hsize\textwidth\columnwidth\hsize\csname @twocolumnfalse\endcsname

\title{Triple sign reversal of Hall effect in HgBa$_{2}$CaCu$_{2}$O$_{6}$ thin
films after heavy-ion irradiations}
\author{W. N. Kang,$^{1,3}$ B. W. Kang,$^{2}$ Q. Y. Chen,$^{3}$ J. Z. Wu,$^{2}$ Y.
Bai,$^{1}$ W. K. Chu,$^{3}$ D. K. Christen,$^{4}$ R. Kerchner,$^{4}$ and
Sung-Ik Lee$^{1}$}
\address{$^{1}$ National Creative Research Initiative Center for Superconductivity,
Department of Physics}
\address{Pohang University of Science and Technology, Kyungbuk 790-784, Korea}
\address{$^{2}$ Department of Physics and Astronomy, University of Kansas, Lawrence,
Kansas 66045, USA}
\address{$^{3}$ Texas Center for Superconductivity, University of Houston, Houston,
Texas 77204, USA}
\address{$^{4}$ Solid State Division, Oak Ridge National Laboratory, Oak Ridge,
Tennessee 37831, USA}
\date{\today }
\draft
\maketitle

\begin{abstract}
The triple sign reversal in the mixed-state Hall effect has been observed
for the first time in ion-irradiated HgBa$_{2}$CaCu$_{2}$O$_{6}$ thin films.
The negative dip at the third sign reversal is more pronounced for higher
fields, which is opposite to the case of the first sign reversal near T$_c$
in most high-T$_c$ superconductors. These observations can be explained by a
recent prediction in which the third sign reversal is attributed to the
energy derivative of the density of states and to a temperature-dependent
function related to the superconducting energy gap. These contributions
prominently appear in cases where the mean free path is significantly
decreased, such as our case of ion-irradiated thin films.

\end{abstract}

\pacs{PACS number:74.60.Ge, 74.25.Fy, 74.72.Gr}

\vskip0.5pc] \newpage

The Hall anomaly in the mixed state of type II superconductors is one of the
most attractive subjects, both experimentally and theoretically, in the
field of vortex dynamics. According to classical theories \cite{Bardeen65},
the vortex motion due to the Lorentz force should generate a Hall voltage
with the same sign as observed in the normal state because normal electrons
in the vortex cores effectively produce this voltage. To the contrary, a
puzzling sign reversal of the Hall effect has been observed in various
conventional superconductors, such as impure Nb \cite{Noto76,Usui68} and V
crystals \cite{Usui68}, and Nb thin films \cite{Hagen90}, and in some high-T$%
_c$ superconductors (HTS), such as YBa$_2$Cu$_3$O$_7$ crystals \cite{Kang96}
and La$_{2-x}$Sr$_x$Cu$_2$O$_4$ \cite{Matsuda95}. Furthermore, a double sign
reversal has been observed in highly anisotropic HTS, such as Bi$_2$Sr$_2$%
CaCu$_2$O$_8$ crystals \cite{Samoilov93}, Tl$_2$Ba$_2$CaCu$_2$O$_8$ films 
\cite{Hagen91}, and HgBa$_{2}$CaCu$_{2}$O$_{6}$ (Hg-1212) films \cite{Kang97}%
. Various models related to two band effects \cite{Hirsch91}, induced
pinning \cite{Wang94}, a superconducting fluctuation \cite{Aronov90}, and a
flux backflow \cite{Hagen90} have been proposed to interpret these Hall
anomalies, but they have not been able to explain the experimental results.
Therefore, the origin of the mixed-state Hall effect still remains an
unsolved subject.

An interesting microscopic approach based on a time-dependent
Ginzburg-Landau theory has been proposed in a number of papers \cite
{Dorsey92,Kopnin93,Otterlo95,Kopnin95}. In this approach, the mixed-state
Hall voltage in type II superconductors is determined by the quasiparticle
and hydrodynamic contributions of the vortex cores. Since the sign of
hydrodynamic term is determined by the energy derivative of the density of
states \cite{Otterlo95,Kopnin95}, if that term is negative, a sign anomaly
can appear. This theory is qualitatively consistent with experimental
results \cite{Kang97,Ginsberg95}, especially for high magnetic fields and
for temperatures near T$_c$.

Recently, Kopnin \cite{Kopnin96} has developed a modified theory which
includes an additional force arising from charge neutrality effects. In this
theory, interestingly, he anticipated the possibility of a third sign
anomaly when the system remains moderately clean, this anomaly would even
occur at low temperatures. This implies that the third sign reversal could
be observed if the mean free path of a system were reduced from the clean
limit of ${\it {l} > \xi}$ to the moderately clean limit of ${\it {l} \sim
\xi}$, where ${\it {l}}$ is the mean free path and $\xi$ is the
superconducting coherence length. This may be the case for columnar defects
due to high-density ion irradiations.

Hg-1212 thin films are suitable candidates for observing the third sign
reversal because the general trend of the negative dip at low field near T$_c
$ still remains at higher fields \cite{Kang99}, a situation which is clearly
different from the cases of the Bi and the Tl compounds \cite
{Samoilov93,Hagen91}. This suggests that at higher fields, the negative
contribution due to the additional transverse force \cite{Kopnin96} in the
Hg-1212 compound is more substantial than that in either the Bi or Tl
compounds.

In this Letter, we present the first report on an observation of triple sign
reversal in superconducting Hg-1212 thin films containing columnar defects
produced by 5-GeV Xe ions. The level of the dose, 1.5 $\times$ 10$^{11}$
ions/cm$^2$, was equivalent to the mean distance, less than 258 $\AA$,
between the columnar defects, thus effectively reducing the mean free path
of the samples, even at low temperatures. Consequently, we were able to
observe, for the first time to the best of our knowledge, the triple sign
reversal predicted by the microscopic theory of nonequilibrium
superconductivity. This observation will provide new insight, we believe,
into the flux dynamics in type II superconductors.

The fabrication process \cite{Yun96,Kang98} and the transport properties 
\cite{Kang99,BWKang97} of the Hg-based thin films used in this study were
previously reported in detail. The mid-transition temperatures T$_c$ of the
as-grown thin films on (001) SrTiO$_3$ substrate were 122 - 124 K. The
critical current density at zero field was $\sim 10^6$ $A/cm^2$ at 100 K.
The X-ray diffraction pattern indicate highly oriented thin films with the c
axis normal to the substrate plane. The minor phase of HgBa$_{2}$Ca$_{2}$Cu$%
_{3}$O$_{8}$ was less than 5 \%. The ion irradiation was performed at the
Superconducting Cyclotron Center at Michigan State University by using 5-GeV
Xe ions. The irradiation was done at room temperature along a direction
normal to the film surface. The irradiation dose was 1.5 x 10$^{11}$ ions/cm$%
^2$, which corresponded to a matching field, B$_\phi$, of $\sim$ 3 T. These
ions produced continuous amorphous tracks with diameters of 50 - 100 $\AA$
in the Hg-1212 thin films. The Hall resistivity $\rho_{xy}$ and the
longitudinal resistivity $\rho_{xx}$ were measured simultaneously using a
two-channel nanovoltmeter (HP34420A) and the standard five-probe dc method.
A magnetic field was applied parallel to the c axis of the Hg-1212 films. $%
\rho_{xy}$ was extracted from the antisymmetric parts of the Hall voltages
measured under opposite fields. The applied current densities were 250 - 500
A/cm$^2$. Both $\rho_{xy}$ and $\rho_{xx}$ were Ohmic at the currents used
in this study.

Typical temperature dependences of $\rho_{xx}$ before (B$_\phi$ = 0 T) and
after heavy-ion irradiation (B$_\phi$ = 3 T) for fields up to 8 T are shown
in Fig. 1. A large enhancement of the zero-resistance temperature, T$%
_{c,zero}$, which is due to strong pinning by the columnar defects, is
clearly visible. This is consistent with the results of previous works \cite
{Kang96,Budhani93} on HTS with columnar defects. We observe that the
enhancement of T$_{c,zero}$ above 3 T is rather small compare to that of
below 3 T, indicating that depinned vortices with a density of $n_{\phi} =
(H - B_{\phi})/\Phi_o$ really contribute to the resistivity, where $\Phi_o$
is the flux quantum. The enhancement of pinning by the columnar defects
becomes effective below the temperature T* which is marked in Fig. 1 by an
arrow.

Figure 2 (a) shows $\rho_{xy}$ before and after irradiation for H = 2, 4, 6,
and 8 T. For the data at H = 2 T, the first sign reversal, which appears in
the vicinity of the transition temperature, does not shift after the
irradiation, while the second sign reversal shifts to higher temperature.
These double sign reversals are not very rare for HTS. For the irradiated
sample, however, if we look at $\rho_{xy}$ on a magnified scale we can
observe a third sign reversal at a relatively lower temperature, as shown in
Fig. 2 (b). The negative dip becomes clear with increasing field, a finding
which is contrary to the one \cite{Kang96,Budhani93} for the first sign
reversal in HTS. For the unirradiated thin films, however, no third sign
change is observed for fields up to 8 T. The inset in Fig. 2 shows the
third-sign-reversal regions where $\rho_{xy}$ is positive (P), negative (N),
and zero (Z), ${\it {i.e.}}$, below the resolution in our experiment. Thus,
we claim that we clearly observed a third sign reversal for the irradiated
thin films.

Now the question arises as to why such a multiple Hall sign reversal is
possible for some superconductors. Is there any relevant explanation for
this phenomenon? Fortunately, a recent study by Kopnin \cite{Kopnin96}
claims that such a phenomenon is possible if the force arising from the
effects of vortex motion on the pairing interaction, which was neglected in
previous works \cite{Dorsey92,Kopnin93,Otterlo95,Kopnin95}, is added to the
Lorentz force. The additional force is induced by the kinetic effect of
charge imbalance relaxation and thus depends on the difference between the
charge densities of the system in the superconducting and the normal states.
According to this theory, the Hall conductivity $\sigma_H$ due to the motion
of a single vortex is determined by three terms: localized excitations,
delocalized excitations, and an additional force term. Therefore, $\sigma_H$
can be expressed by the sum of three terms:

\begin{equation}
\sigma_H=\sigma_{H}^{(L)}+\sigma_{H}^{(D)}+\sigma_{H}^{(A)}.
\end{equation}
Using $\hbar = c = k_B = 1$, the Hall conductivity due to localized
excitations,$\sigma_{H}^{(L)}$, is given by

\begin{equation}
\sigma_{H}^{(L)} \sim \frac{Ne}{B}\frac{(\omega_o\tau)^2}{1+(\omega_o\tau)^2}%
,
\end{equation}
where N is the density of carriers, $\tau$ is the relaxation time, $\omega_o
\sim \Delta^2/E_F$ is the distance between the energy levels in the vortex
core, and E$_F$ is the Fermi energy. The portion of Hall conductivity
contributed by the additional force,$\sigma_{H}^{(A)}$, is

\begin{equation}
\sigma_{H}^{(A)} \sim \frac{e}{B\lambda} \left( \frac{\partial\nu}{%
\partial\zeta} \right) \Delta^{2} \beta(T),
\end{equation}
where $\lambda$ is the BCS coupling strength, ${\partial\nu}/{\partial\zeta}$
is the energy derivative of the density of states, $\Delta$ is the
superconducting energy gap, and $\beta(T)$ is a temperature-dependent
function and is positive, which depends on temperatures in the following
way: either $\beta$ = 1 near T$_c$ or $\beta(T) \sim \Delta/[Tln(\Delta/T)]$
at low temperatures \cite{Kopnin96}. Since the delocalized excitation term, $%
\sigma_{H}^{(D)}$, is due to the density of quasiparticles outside the
vortex core, the sign of $\sigma_{H}^{(D)}$ is the same as the sign of the
normal-state Hall conductivity and $\sigma_{H}^{(D)}$ is very small at low
temperatures compared to $\sigma_{H}^{(L)}$ \cite{Kopnin95}. Due to this, we
will simply neglect $\sigma_{H}^{(D)}$ at low temperatures. It is found \cite
{Harris94} that the tangent of the Hall angle, $tan\Theta \sim \omega_o\tau$%
, is very small $(\sim 0.01)$ for the dirty region near T$_c$ and approaches 
$\sim 1$ for the superclean region at T $\ll T_c$. This is consistent with
the theoretical calculation \cite{Kopnin95}. $\sigma_{H}^{(A)}$ deduced from
charge imbalance relaxation is determined by the energy derivative of the
density of states ${\partial\nu}/{\partial\zeta}$ at the Fermi surface and
by $\beta(T)$. Since this latter term is negative, the sign of $\sigma_{H}$
can critically depend on this term. If $\sigma_{H}^{(A)}$ is negative and if
it is the dominant contribution, then $\sigma_{H}$ can be negative.

In order to estimate the sign of ${\partial\nu}/{\partial\zeta}$, we had
better comment on the symmetry of the superconducting order parameters. Very
recently, Himeda ${\it {et}}$ ${\it {al.}}$ \cite{Himeda97} calculated the
microscopic structure of the vortex cores in HTS by using the
two-dimensional t-J model for a wide range of doping rates. They argued that
the density of states split into two levels due to mixing of the s- and the
d-wave components in the underdoped regions. The typical density of states
for d-wave superconductors was observed in the overdoped regions. This
indicates that the sign of ${\partial\nu}/{\partial\zeta}$ in $%
\sigma_{H}^{(A)}$ should not be based on the BCS s-wave theory \cite
{Nagaoka98}. Within the context of the Kopnin's theory, however, we can
estimate the sign of $\sigma_{H}^{(A)}$ from the experimental results. From
the data for H = 2 T in Fig. 2, for example, the sign of the additional
force term is negative simply because $\sigma_{H}$ is positive and $%
\sigma_{H}^{(A)}$ is negative; thus, ${\partial\nu}/{\partial\zeta}$ is
negative.

According to Eq. (1), $\sigma_{H}$ allows multiple sign reversals as a
function of temperature and mean free path. The sign reversals arise from
competition between a positive $\sigma_{H}^{(L)}$ and a negative $%
\sigma_{H}^{(A)}$. For the dirty case with ${\it {l} < \xi}$, $%
\sigma_{H}^{(A)}$ is dominant because $(\omega_o\tau)^2 \sim 10^{-4}$ is
very small as T $\rightarrow$ T$_c$. Thus, the sign of $\sigma_{H}$ can be negative. For
the clean case with ${\it {l} > \xi}$, $\sigma_{H}^{(L)}$ is dominant
because the magnitude of $(\omega_o\tau)^2$ is very large compared to its
magnitude for the dirty case; thus $\sigma_{H}$ is positive in the
low-temperature region. This is a plausible interpretation for the double
sign reversal observed in Fig. 2. The double sign reversal was also observed
in highly anisotropic HTS, such as Bi- and Tl-based compounds \cite
{Samoilov93,Hagen91}.

Now, the problem is to explain the triple sign reversal. The first thing we
should notice is that the triple sign reversal is observed only in the
ion-irradiated samples. In that case, $\sigma_{H}^{(L)}$ decreases seriously
because $\omega_o\tau$ is reduced drastically by a change from the clean to
the moderately clean or the dirty case. Then, a third sign reversal in the
mixed state is quite natural, and this interpretation is in good agreement
with the experimental observations. To explain this in detail, we should
point out that at low temperatures, $\sigma_{H}^{(L)}$ and $\sigma_{H}^{(A)}$
have different temperature dependences through $\omega_o\tau$ and $\beta(T)$%
. Since $\beta(T)$ increases as $\sim$ 1/T with decreasing temperature and $%
(\omega_o\tau)^2$ is still small for the moderately clean region, $%
\sigma_{H}^{(A)}$ can exceed $\sigma_{H}^{(L)}$ again at low temperatures 
\cite{Kopnin96}, especially for the moderately clean case. As a result, we
should expect a third sign reversal of the mixed-state Hall effect if
high-density impurities exist, which is in agreement with our observation in
Fig. 2.

At this point, it is meaningful to compare the temperature dependence of the
Hall angles before and after ion irradiation, as shown in Fig. 3. As the
temperature decreases from the normal state to the superconducting state, $%
tan{\Theta}$ of the pristine sample increases steeply and then shows a peak
at a relatively low temperature. The maximum magnitude of $tan{\Theta}$ at H
= 8 T is much larger than that observed in YBCO crystals \cite{Harris94}
which are believed to be very clean superconductors. Note that $tan{\Theta}$
is reduced significantly by the ion irradiation, even above T$^*$ where the
pinning is not important. This result should be explained by an impurity
effect rather than by the pinning effect. This strongly supports the above
interpretation that $\sigma_{H}^{(L)}$ can decrease if a clean system become
a moderately clean after the irradiation.

Note that we observe the third sign reversal for Hg-1212 thin films
irradiated with an ion dose of 1.5 x 10$^{11}$ ions/cm$^2$, which
corresponds to an average distance of 258 $\AA$ between the columnar
defects. If we consider columnar defects with the diameter 50 - 100 $\AA$
and universal defects, such as oxygen vacancies, in the pristine thin films,
the mean free path is very low and is much smaller than the value reported
for YBCO crystals in the low-temperature region \cite{Krishana95}.
Therefore, we can observe a triple sign reversal after heavy-ion
irradiations because the irradiated thin films are probably moderately
clean, even at low temperatures.

As a final note, since the above interpretation is based on the assumption
that there are localized states at the core level, it is worth mentioning
the existence of localized core states in d- wave superconductors. Localized
core states, which are consistent with the predictions of theoretical works 
\cite{Morita97}, have been observed in HTS by using various experimental
probes, such as a far-infrared spectroscopy \cite{Karrai92} and scanning
tunneling spectroscopy \cite{Maggio95}. Furthermore, in the moderately clean
case, Kopnin and Volovik \cite{Kopnin98} have observed that $\sigma_{H}^{(L)}
$ for d-wave superconductors is similar to the previous result \cite
{Kopnin95} based on s-wave superconductors. On the other hand, in a recent
calculation \cite{Franz98} for the quasiparticle state in a d-wave
superconductor, Frantz and Tesanovic have claimed that the bound state in
the vortex core was not observed.

In summary, the Hall effect in Hg-1212 films has been studied before and
after irradiation by high-energy Xe ions. After irradiation with a dose of
1.5 x 10$^{11}$ ions/cm$^2$, we find that columnar defects play an important
role not only as strong pinning sites but also as high-density impurities
which can effectively reduce the mean free path even at low temperatures. As
a result, we observe a triple sign reversal, which can be qualitatively
interpreted within the framework of a recent model based on the
nonequilibrium microscopic theory.

This work is partially supported by the Ministry of Science and Technology
of Korea through the Creative Research Initiative Program. The work at
University of Kansas is supported by AFOSR, NSF, NSF EPSCoR, and DEPSCoR.

\begin{figure}[tbp]
\caption{$\rho_{xx}$ vs T curves for pristine (B$_\phi$ = 0 T, open circles)
and irradiated (B$_\phi$ = 3 T, solid lines ) Hg-1212 films in magnetic
fields of 2, 4, and 8 T. The enhancement of pinning by columnar defects
becomes effective below the temperature T$^*$ which is marked by the arrow.}
\end{figure}

\begin{figure}[tbp]
\caption{$\rho_{xy}$ vs T for pristine (B$_\phi$ = 0 T, open circles) and
irradiated (B$_\phi$ = 3 T, solid lines ) Hg- 1212 films. (b) Corresponding
plot of $\rho_{xy}$ on a magnified scale so that the triple sign reversal
can be observed. The inset shows the third-sign-reversal regions where $%
\rho_{xy}$ for irradiated thin films is positive (P), negative (N), and zero
(Z).}
\end{figure}

\begin{figure}[tbp]
\caption{$tan{\Theta}$ vs T for pristine (B$_\phi$ = 0
T, open circles) and irradiated (B$_\phi$ = 3 T, solid lines) Hg-1212 films.
The pinning by the columnar defects is negligible above T$^*$, as shown in
Fig. 1.}
\end{figure}

\end{document}